\documentclass[11pt,a4paper]{article}

\usepackage[T1]{fontenc}
\usepackage[utf8]{inputenc}
\usepackage[scaled]{helvet}

\usepackage{sansmath}\sansmath
\DeclareSymbolFont{sfgreek}{OT1}{cmss}{m}{n}
\DeclareMathSymbol{\Gamma}{\mathord}{sfgreek}{"00}
\DeclareMathSymbol{\Delta}{\mathord}{sfgreek}{"01}
\DeclareMathSymbol{\Theta}{\mathord}{sfgreek}{"02}
\DeclareMathSymbol{\Lambda}{\mathord}{sfgreek}{"03}
\DeclareMathSymbol{\Xi}{\mathord}{sfgreek}{"04}
\DeclareMathSymbol{\Pi}{\mathord}{sfgreek}{"05}
\DeclareMathSymbol{\Sigma}{\mathord}{sfgreek}{"06}
\DeclareMathSymbol{\Upsilon}{\mathord}{sfgreek}{"07}
\DeclareMathSymbol{\Phi}{\mathord}{sfgreek}{"08}
\DeclareMathSymbol{\Psi}{\mathord}{sfgreek}{"09}
\DeclareMathSymbol{\Omega}{\mathord}{sfgreek}{"0A}
\usepackage[margin=2.2cm]{geometry}
\usepackage{amsmath,amssymb}
\usepackage{booktabs,array,tabularx}
\usepackage{graphicx}
\usepackage{xcolor}
\usepackage[labelfont=bf,labelsep=period,font=small]{caption}
\usepackage[numbers,sort&compress]{natbib}
\usepackage{microtype}
\usepackage{setspace}
\usepackage{hyperref}
\hypersetup{colorlinks=true,linkcolor=black,citecolor=black,urlcolor=blue}
\usepackage{titlesec}
\titleformat{\section}{\normalfont\Large\bfseries}{}{0pt}{}
\titleformat{\subsection}{\normalfont\large\bfseries}{}{0pt}{}
\titlespacing*{\section}{0pt}{1.6em}{0.6em}
\titlespacing*{\subsection}{0pt}{1.2em}{0.4em}

\begin{document}

\begin{center}
{\LARGE\bfseries Photonic processors that carry their own\\[2pt] unclonable fingerprint\par}
\vspace{1.0em}
{\large
Oxel Arnal Martínez$^{a,*}$,
Pablo Martínez-Carrasco Romero$^{a,*}$, David Álvarez Outerelo$^{b}$,
Francisco Javier Fraile-Peláez$^{b}$, 
and José Capmany Francoy$^{a}$\par}
\vspace{0.8em}
{\small
$^{a}$iTEAM Research Institute, Universitat Politècnica de València, Valencia, 46022, Spain\\
$^{b}$Dept. Teoría de la Señal y Comunicaciones, Universidad de Vigo, E.I. Telecomunicación,\\
Campus Universitario, E-36202 Vigo (Pontevedra), Spain\\[0.4em]
$^{*}$Corresponding authors: \href{mailto:oarnmar@iteam.upv.es}{oarnmar@iteam.upv.es}, \href{mailto:pmarrom@iteam.upv.es}{pmarrom@iteam.upv.es}\par}
\end{center}

\vspace{0.6em}
\noindent\rule{\linewidth}{0.4pt}
\vspace{-0.2em}

{\bfseries Abstract}

Physically unclonable functions (PUFs) derive cryptographic keys from uncontrollable fabrication disorder, removing the need to store secrets in vulnerable non-volatile memory. Programmable photonic processors are natural candidates for PUFs, because each of their many tunable interferometers carries its own fabrication imperfections. However, we show that the mesh topology itself funnels light along short paths, so that most of this disorder contributes little to the response. Here we introduce an isotropic ring routing scheme that steers light to the core of a hexagonal mesh and redistributes it uniformly towards the perimeter, suppressing a distance-dependent power decay of 14.7~dB and raising the mean Shannon entropy of the response from 1.95 to 3.48 bits by reprogramming alone. Experiments on two independently fabricated 72-unit silicon processors show that a nominally identical device cannot reproduce the target's responses even after its passive phase profile is fully compensated. Combining isotropic routing with spectral readout yields 1107 bits per challenge, exceeding previous results by two orders of magnitude, with 49.8\% inter-challenge and 50.0\% inter-chip Hamming distances and a 4.9\% intra-chip distance. These results establish general-purpose programmable photonic processors as reconfigurable, high-capacity hardware roots of trust.

\textbf{Keywords:} programmable photonics, optical signal processing, physically unclonable function

\vspace{-0.4em}
\noindent\rule{\linewidth}{0.4pt}

\section{Introduction}

The rapid expansion of the Internet of Things (IoT), AI-driven computing, and massive data transmission has significantly increased the demand for robust and scalable hardware security \cite{tomkos2020,wang2020}. Conventional cryptographic systems rely on secret keys stored in nonvolatile memory, which leaves them exposed to physical tampering, invasive key extraction and increasingly sophisticated attacks \cite{anderson1996,zhou2005,agrawal2003}. Physically unclonable functions (PUFs) offer a fundamentally different approach to authentication and key generation: rather than storing a secret, they derive it on demand from the physical structure of the device itself \cite{pappu2002}.

PUFs exploit the uncontrollable randomness introduced during manufacturing to generate a unique hardware fingerprint that acts as a root of trust. A PUF maps an input stimulus to a device-specific output, forming a challenge-response pair (CRP) that is exceedingly difficult to predict or reproduce without physical access to the original device. By deriving security from intrinsic physical characteristics rather than stored secret keys, PUFs are inherently resistant to cloning and to invasive attacks, because probing the device tends to alter the very structure that defines its response \cite{pappu2002,herder2014,gao2020,bohm2012,almeer2023}.

To date, PUF research has been dominated by electronic implementations, particularly on widely available FPGA platforms, with prominent architectures including ring oscillators, static random-access memory and arbiter PUFs \cite{gassend2002,suh2007,lee2004}. Although these electronic PUFs offer excellent integration, robustness, and unclonability, they remain vulnerable to non-invasive modelling, side-channel and machine-learning attacks, because their challenge-response behaviour can often be learned from a limited set of observed CRPs \cite{ruhrmair2010,ruhrmair2013}. These limitations have sparked growing interest in optical PUFs, which exploit the richer physics, analog signal propagation, and higher information capacity of photonic systems to generate more complex responses \cite{pavanello2021,gandla2025,mannandrews2025}.

Optical PUFs have been demonstrated in a wide range of platforms, including free-space scattering systems \cite{nocentini2024,wang2024}, multimode-fiber PUFs \cite{mesaritakis2018}, and integrated wavelength-dependent photonic structures such as chaotic microcavities \cite{bosworth2020,grubel2017,chang2026}, Moiré crystals \cite{tarik2022} and Bragg gratings \cite{mahdian2024}. These architectures, however, face important limitations. Free-space and fiber implementations are often bulky and lack compatibility with CMOS technology, limiting their integration with the digital electronic circuitry required for cryptographic key management. Meanwhile, integrated photonic PUFs are typically constrained in their key-generation capability. Once the device has been fabricated, the space of possible CRPs is largely fixed, as the physical characteristics that determine its behavior cannot be modified. Consequently, the number of unique keys that can be generated is constrained by the device's static structure, limiting its scalability and long-term flexibility.

\begin{figure}[t]
\centering
\includegraphics[width=\linewidth]{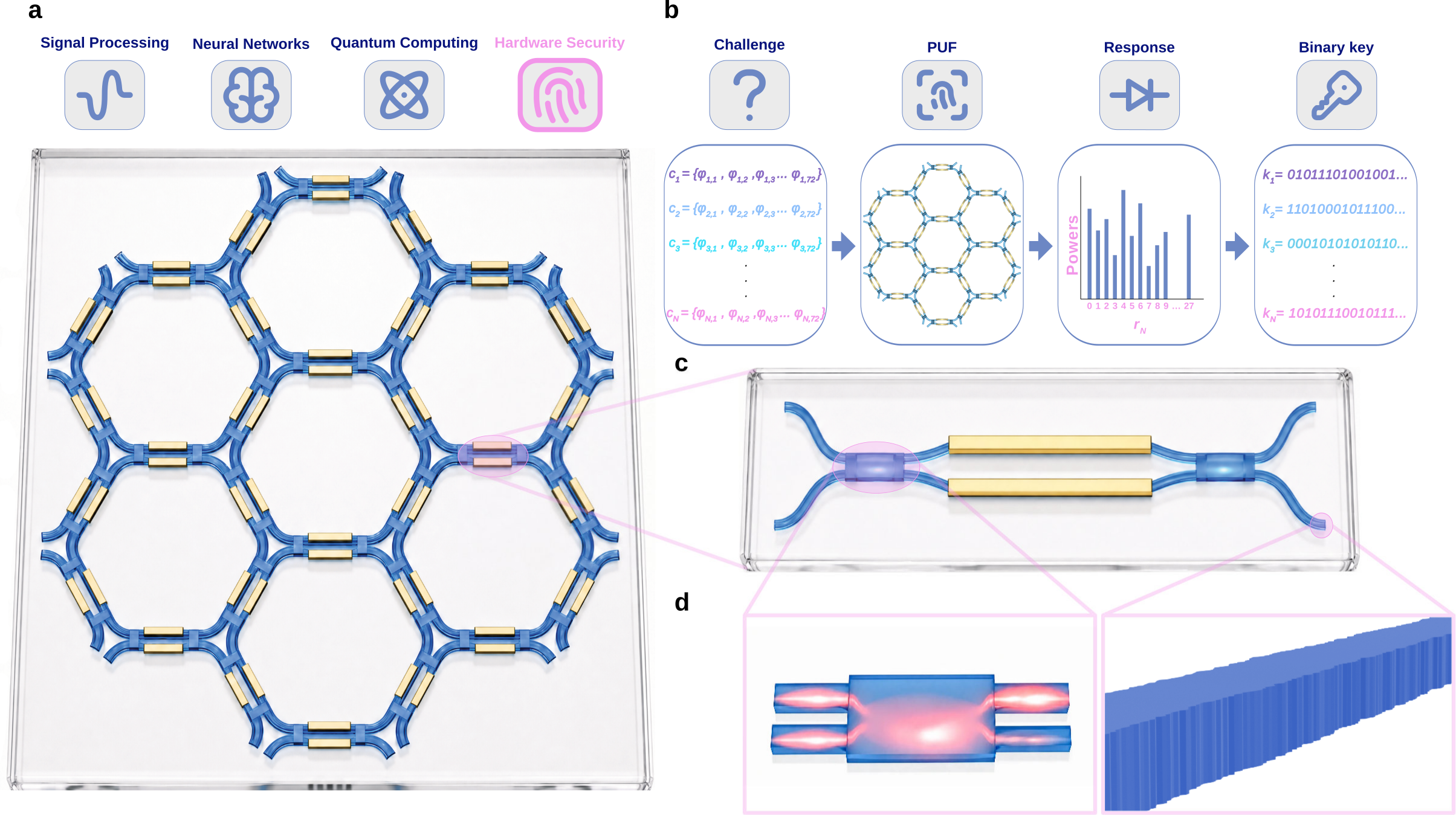}
\caption{\textbf{Programmable photonic circuits for PUF implementation.} (a) Schematic of a programmable hexagonal photonic mesh for multipurpose applications. (b) General implementation of a PUF in a programmable photonic circuit, where a set of phase shifts (challenge) is applied to the MZIs, and the outgoing optical signals from the outputs are photodetected and converted into a unique binary key. (c) Inset view of an individual MZI with its characteristic deviations. (d) Detailed view of the MMIs and waveguides, showing fabrication variations that result in imperfect beam splitting, a specific passive offset, and optical loss.}
\label{fig:1}
\end{figure}

Programmable photonic integrated circuits (PICs) based on hexagonal meshes of interconnected tunable Mach--Zehnder interferometers (MZIs), also referred to as tunable basic units (TBUs), have emerged as a promising platform for optical PUFs \cite{bogaerts2020,perezlopez2020,zhuang2015}. As illustrated in Figure~\ref{fig:1}, the unique fingerprint of each device arises from unavoidable nanoscale fabrication variations affecting every building block of the photonic mesh. These variations introduce coupling and loss deviations in the multimode interference (MMI) couplers of the MZIs, passive phase offsets due to waveguide sidewall roughness, and fluctuations in the responsivity and dark current of the output photodetectors \cite{xing2023,mikkelsen2014}. Collectively, these intrinsic imperfections generate a unique hardware signature for each device, making programmable photonic meshes an attractive platform for cryptographic key generation while preserving their multifunctional capabilities for optical signal processing and computing, so that the same chip can act both as a processor and as its own hardware root of trust.

Previous PUF demonstrations in programmable meshes, however, have either confined key extraction to small, locally synthesized sub-circuits \cite{zhu2026,tzekas2026} or relied on additional resources, such as neural-network processing \cite{sarantoglou2025} or random projection \cite{sarantoglou2026} to compensate for the limited entropy of the raw optical response. Why the raw response of a large, disordered mesh carries so little entropy has not been established, and it remains unclear how much of the fabrication disorder distributed across the mesh actually reaches its outputs. Maximizing the intrinsic entropy of the mesh would turn such additional resources from a necessity into an optional enhancement.

Here, we identify the topology of the mesh as the fundamental bottleneck for entropy extraction and show that it can be removed by optical routing alone. Under unconstrained routing, light remains concentrated along short paths near the input, so that distant outputs carry little power and therefore little information. We introduce an Isotropic Ring routing topology that steers the light to the core of the mesh and redistributes it uniformly towards the perimeter, eliminating the distance-dependent power decay and increasing the mean Shannon entropy of the response by 78\%. Experiments on two independently fabricated processors confirm strong device-to-device uniqueness, which persists even when the passive phase profile of the target device is fully compensated in a nominally identical device. Finally, by combining the Isotropic Ring with a spectral readout that exploits the wavelength response of every output port, we generate 1107 bits per challenge with near-ideal uniqueness and without any post-processing: more than two orders of magnitude more than previous mesh PUFs that use none, and more than four times the key length of those that rely on it for key expansion and randomization. These results bridge the high entropy density of scattering-based PUFs with the scalability and CMOS compatibility of integrated photonics, and establish general-purpose programmable photonic processors as practical, reconfigurable hardware roots of trust.

\section{Results}

\subsection{Topology and entropy enhancement}

The primary source of entropy in a programmable photonic mesh arises from the random phase offsets of its constituent MZIs, which are approximately uniformly distributed over $[0, 2\pi)$. However, when these random phases propagate through the constrained topology of the mesh, their collective effect resembles a random walk: although each individual step is unbiased, the probability of reaching distant states decreases as the number of steps increases. Analogously, optical power progressively concentrates towards output ports connected through shorter and more probable paths, while ports requiring longer propagation paths receive substantially weaker signals. Consequently, the intrinsic randomness of the MZI phases does not translate directly into an unbiased response space; instead, the mesh topology transforms this randomness into a structured and predictable distribution.

\begin{figure}[t]
\centering
\includegraphics[width=\linewidth]{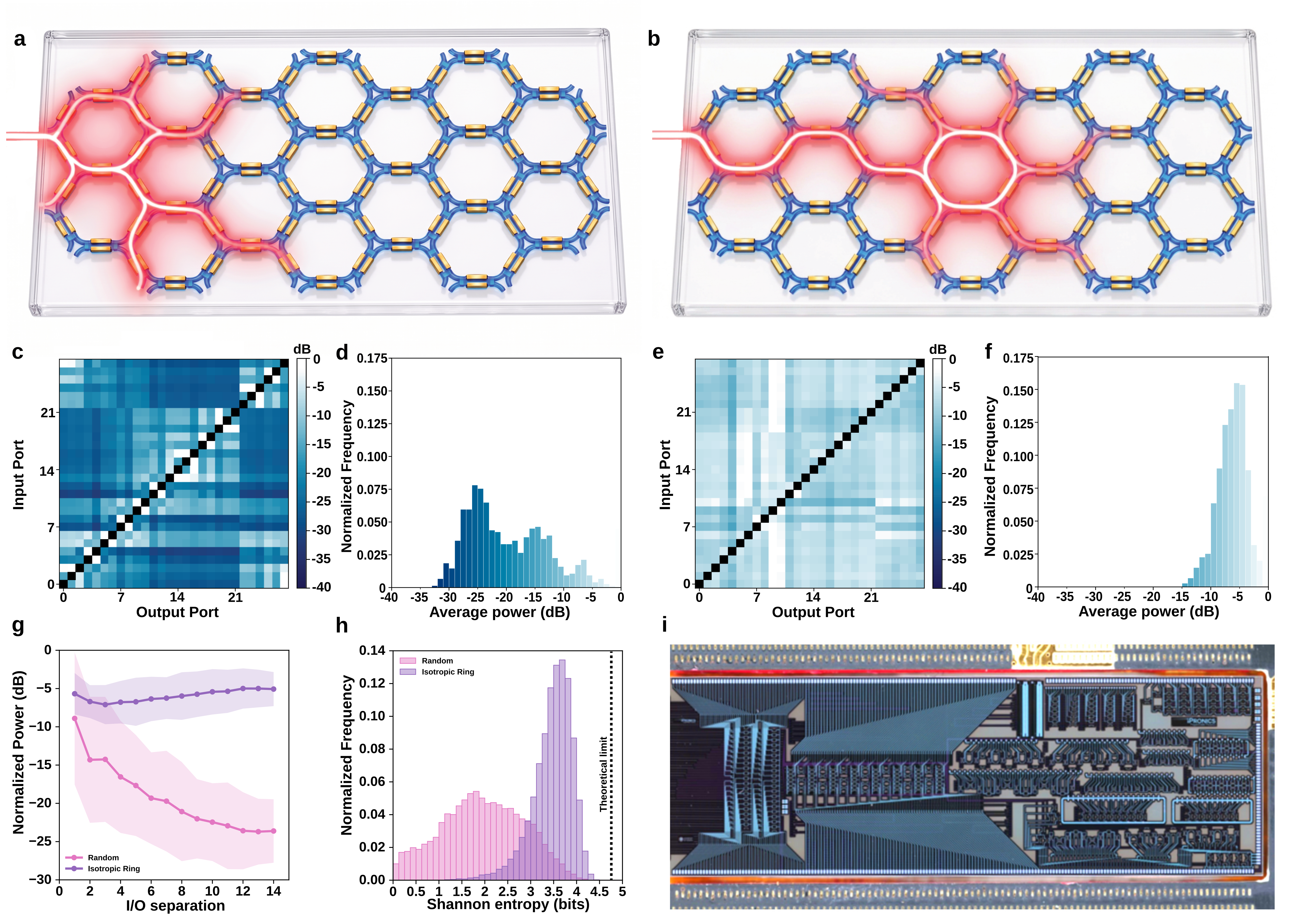}
\caption{\textbf{Routing topologies and experimental power response of the photonic mesh for the random baseline and the Isotropic Ring.} (a, b) Schematics of both configurations for a representative input port. (c, e) Input-output power maps, and (d, f) histograms of the average output power, for the random (c, d) and Isotropic Ring (e, f) configurations. (g) Normalized output power as a function of input-output separation in number of ports for the random and isotropic ring configurations. The line shows the mean and the shaded band $\pm 1$ standard deviation over 1000 challenges across all the input ports. (h) Histograms of the Shannon entropy computed for each individual challenge for the random and isotropic ring configurations; the dashed line marks the theoretical limit $\log_2(27) = 4.75$ bits. (i) Image of the silicon photonic integrated circuit inside the processor, containing the hexagonal mesh of 72 TBUs.}
\label{fig:2}
\end{figure}

This bias limits the entropy available for PUF operation, however rich the underlying fabrication disorder may be. We address this at its root. In the random baseline (Fig.~\ref{fig:2}a), all 72 TBUs receive random phases and no path is imposed, so the light tends to escape through the shorter, more predictable paths near the input port. In the Isotropic Ring topology (Fig.~\ref{fig:2}b), instead, a fixed subset of TBUs forces the injected light along a defined path from the input port to the mesh core. There, the light is split equally (diffused) across the six branches of the central hexagon and recirculates symmetrically outward, reaching all the ports at the perimeter rather than escaping through the nearest ones. By sacrificing the entropy contribution of this small, fixed subset (8 to 12 of the 72 TBUs; see Methods), the Isotropic Ring routes the signal through a substantially larger number of dynamic MZIs than the random baseline, producing a net gain in the entropy available in the response.

For the experimental demonstration, we used a Smartlight processor (iPronics Programmable Photonics, Fig.~\ref{fig:2}i), that contains a silicon-on-insulator hexagonal mesh of 72 TBUs with 28 optical ports along its perimeter, driven by a continuous-wave laser operating at 1550~nm (see Methods). The state of each MZI depends on its total phase, $\varphi_{\text{total}} = \varphi_{0} + \varphi_{\text{active}}$, which is the sum of its passive phase ($\varphi_{0}$), set by fabrication, and the applied phase ($\varphi_{\text{active}}$) of the specific configuration. A challenge is therefore defined by the input port and the phase configuration of the mesh, with the experimental evaluation comprising 1000 random sets of applied phases drawn uniformly over $[0,2\pi)$. For each challenge, light was injected into the selected input port and the photodetected power was measured at the remaining 27 ports. This generates an analog power vector $\mathbf{P} \in \mathbb{R}^{N}$, normalized to the total received power, where $N = 27$ represents the number of monitored output ports, located along the perimeter of the mesh and ordered by their position on it (see Supplementary Note~\ref{sn:ports}). The procedure was repeated for the random baseline, in which all 72 TBUs are set by the challenge, and for the Isotropic Ring topology.

Figures~\ref{fig:2}c and \ref{fig:2}e show the input-output power maps of the random and Isotropic Ring configurations, and Figures~\ref{fig:2}d and \ref{fig:2}f the corresponding distributions of the average output power. Each entry is the normalized output power averaged over all challenges for the different input ports. In an unbiased mesh, averaging power across challenges would yield a uniform response, with no privileged paths causing input-dependent attenuation. However, the random configuration exhibits clear spatial bias and a heterogeneous distribution, concentrating most of the optical power near the input port and creating pronounced attenuation valleys at more distant outputs. In contrast, the Isotropic Ring topology achieves high input-output power uniformity, resulting in a visibly more homogeneous map and a substantially narrower power distribution. This behavior is evident in Figure~\ref{fig:2}g, which depicts the dependence of normalized output power on input-output port separation, where a separation of 1 represents adjacent ports and 14 corresponds to diametrically opposed ports across the mesh. The random configuration exhibits a pronounced power decay with distance, consistent with the shorter optical paths favored by unconstrained routing, where the mean normalized power drops from $-8.9$~dB at adjacent ports to $-23.6$~dB at diametrically opposed ports. Conversely, the Isotropic Ring effectively eliminates this spatial dependence, with the mean normalized power around $-5.6$~dB changing by only 2~dB over the same range of separations. Finally, to illustrate how this unbiased distribution translates into the information content of the mesh, we calculated the Shannon entropy of each CRP from the power levels across all output ports (see Methods). The resulting entropy distributions are presented in the histograms of Figure~\ref{fig:2}h. The random configuration yields a widely scattered entropy distribution with a mean of 1.95 bits (standard deviation of 0.89 bits), remaining far below the theoretical limit of $\log_2(27) = 4.75$ bits. In contrast, the Isotropic Ring shifts the average significantly higher to 3.48 bits while halving the standard deviation to 0.43 bits. This results in a 78\% increase in the information carried by each response obtained solely by reprogramming the mesh, without any hardware modification.

\subsection{Binarization and digital response}

Converting the analog power vector into a digital key requires robustness against laser intensity fluctuations and global thermal drifts, to which threshold-based binarization is particularly susceptible. We therefore use a spatial differential operator based on the relative power ranking of the outputs, selected after benchmarking alternative strategies (see Supplementary Note~\ref{sn:binarization}).

The key is generated by pairwise comparison of output powers. Because it relies only on relative power levels, this scheme cancels common-mode variations, including laser intensity fluctuations and global coupling losses. Each bit $b_{n}$ is the result of comparing the $n$-th and $(n + d)$-th monitored outputs, so that $b_{n} = 1$ if $P_{n} > P_{n + d}$ and $b_{n} = 0$ otherwise, where the index $d$ dictates the port separation. The output indices are taken cyclically along the perimeter, so that counting continues from the first output after the last, and 26 comparisons are kept for every $d$ so that all keys have the same length. Specifically, $d = 1$ or $d = 26$ evaluates adjacent ports, whereas $d = 13$ targets diametrically opposed ports.

\begin{figure}[t]
\centering
\includegraphics[width=\linewidth]{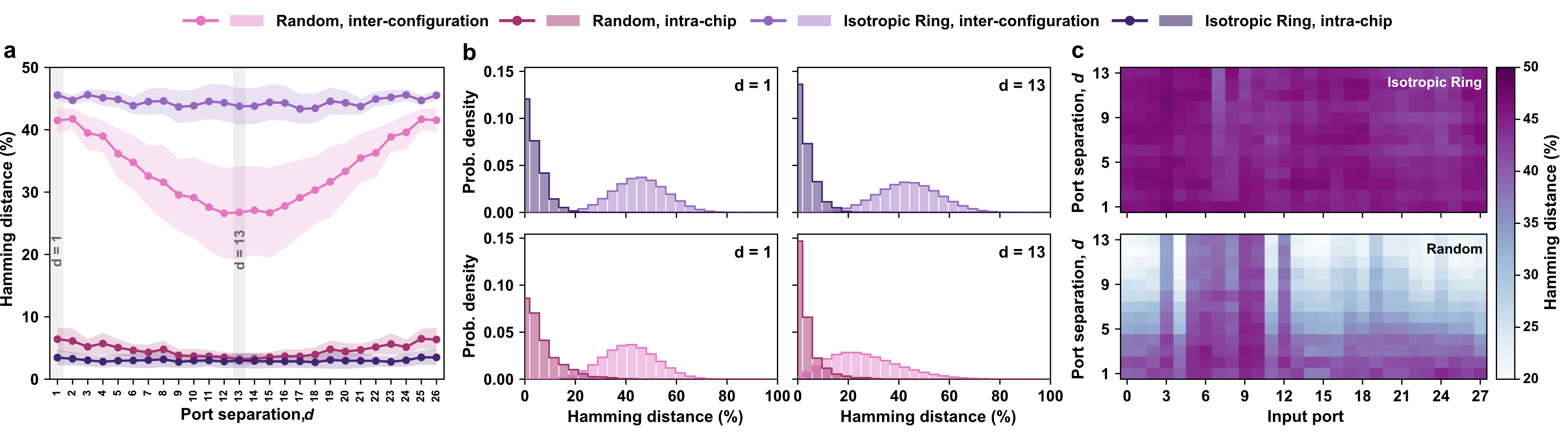}
\caption{\textbf{Binary key extraction performance for Isotropic Ring and random mesh topologies.} (a) Mean inter-configuration and intra-chip Hamming distances as a function of port separation. Shaded bands represent $\pm 1$ standard deviation computed over 1000 challenges across all input combinations. (b) Distributions of inter-configuration and intra-chip HDs for both topologies under minimal separation ($d = 1$, top) and maximum spatial separation ($d = 13$, bottom). (c) Mean inter-configuration HD as a function of the input port and of the port separation $d$, for the Isotropic Ring (top) and the random configuration (bottom). In the random configuration, input ports 3, 5 to 10 and 12 keep an inter-configuration HD closer to the ideal value for all separations, due to their position along the long edge of the mesh.}
\label{fig:3}
\end{figure}

Figure~\ref{fig:3} illustrates the binary key extraction performance across both topologies, using intra- and inter-class Hamming distances (HDs) to quantify the stability and uniqueness of 26-bit keys evaluated across 1000 distinct configurations and all input ports (see Methods). Fig.~\ref{fig:3}a plots the mean HD across mesh configurations as a function of port separation. The Isotropic Ring topology extracts consistently robust keys regardless of port separation, maintaining a stable inter-HD near 45\% and an intra-HD near 3\%. Conversely, the random topology exhibits a strong dependence on spatial separation, with inter-HD dropping from 41\% at adjacent ports to 27\% at distant ones and an intra-HD varying between 6\% and 3\%.

Figure~\ref{fig:3}b shows the HD distributions for both topologies at the two extreme values of $d$ (adjacent ports and maximum spatial separation). At minimal separation ($d = 1$), adjacent ports receive comparable powers, so that bit values are governed by the phase configuration and yield highly unpredictable keys. In both topologies, this behavior manifests as narrow HD distributions centered close to the ideal theoretical value of 50\%. As $d$ increases, path-length asymmetries cause the port closer to the input to capture more power deterministically.

Spatial bias therefore undermines port pairing as an additional degree of freedom for key extraction, and does so in two distinct ways. First, at the output level, pairing distant ports in biased meshes deterministically favors specific channels, resulting in predictable keys and a reduced Hamming distance. While spatial degradation broadens the HD distributions across both topologies, the Isotropic Ring demonstrates greater resilience by maintaining a stable inter-HD near optimal values, whereas the random topology undergoes a pronounced drift toward degraded performance. Second, input port selection adds a geometry-induced bias due to the chip's rectangular footprint as illustrated in Fig.~\ref{fig:3}c across all port separations. Injection through ports along the longer edge promotes a more balanced power distribution, naturally mitigating spatial bias. As a result, the random topology exhibits HD metrics closer to the ideal values for these specific inputs (ports 3 to 12; see Supplementary Note~\ref{sn:ports}), regardless of output port pairing. In contrast, the remaining input ports in the random topology display high sensitivity to output selection, where the mean inter-HD drops to 30.8\%. The Isotropic Ring eliminates these spatial dependencies, achieving a higher inter-challenge HD that is insensitive to port pairing, lower intra-chip HD, and uniform behavior across all input ports.

\subsection{Uniqueness across devices}

\begin{figure}[t]
\centering
\includegraphics[width=\linewidth]{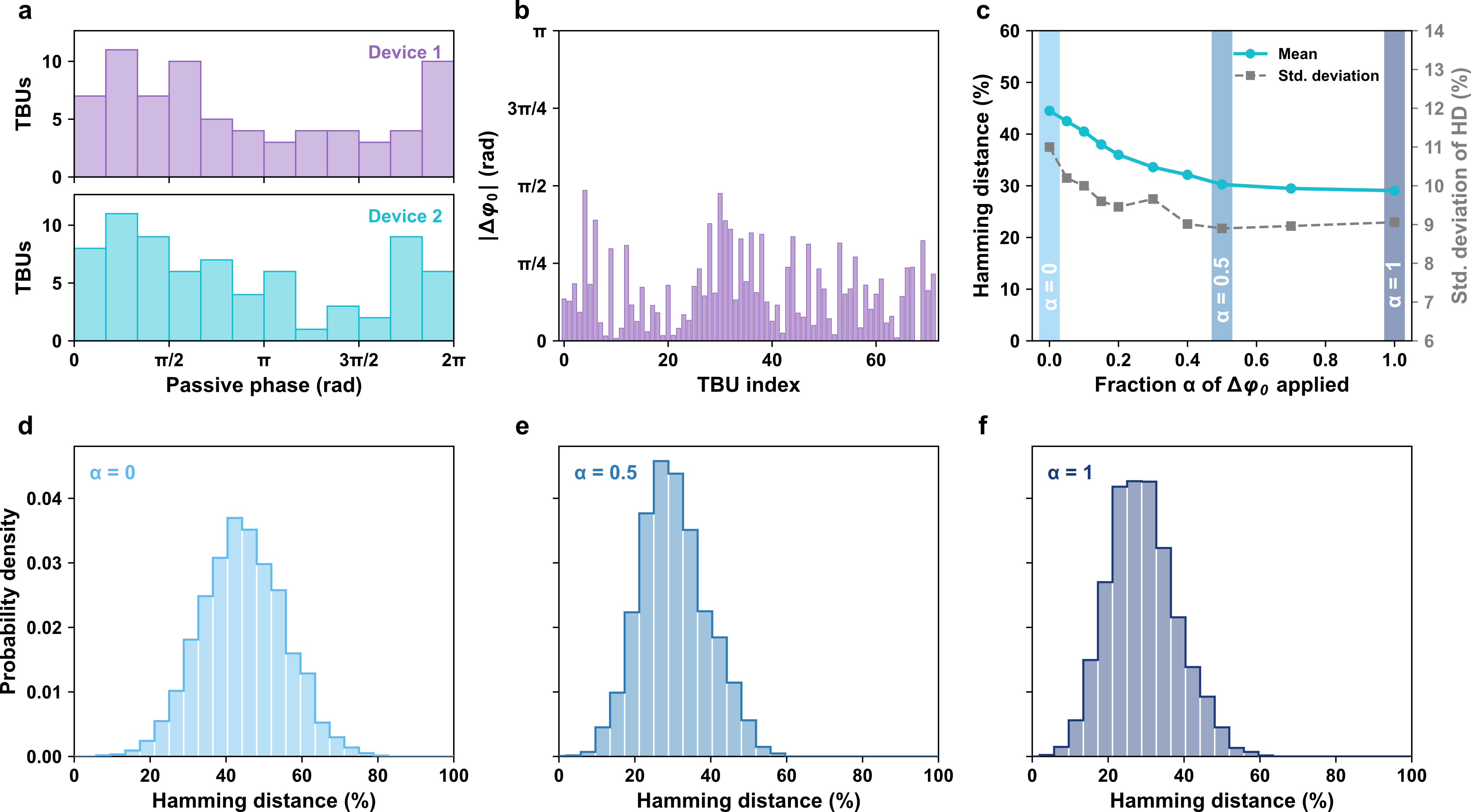}
\caption{\textbf{Uniqueness across devices.} (a) Distribution of the passive phases $\varphi_{0}$ of the 72 TBUs of Device 1 (top) and Device 2 (bottom). (b) Absolute passive phase difference between both devices, $|\Delta\varphi_{0}|$, for each TBU. (c) Mean HD (left axis) and standard deviation (right axis) between the response of Device 2 to the original challenge and the response of Device 1 when a fraction $\alpha$ of $\Delta\varphi_{0}$ is added to its dynamic TBUs. The shaded bands mark the values of $\alpha$ whose distributions are shown in panels d to f. (d) Distribution of the inter-chip HD for identical challenges ($\alpha = 0$). (e, f) Distribution of the HD between both devices for $\alpha = 0.5$ (e) and $\alpha = 1$ (f), corresponding to half and full compensation of the passive phase difference.}
\label{fig:4}
\end{figure}

To assess inter-device uniqueness, we characterized a second Smartlight processor of identical design under the same challenge set and computed the normalized Hamming distance between the resulting 26-bit responses of both chips. As a first step, we measured the passive phase $\varphi_{0}$ of each of the 72 TBUs in both devices. In each device, the passive phases are spread over the whole $[0,2\pi)$ range without a dominant value (Fig.~\ref{fig:4}a), as expected from random fabrication variations. However, corresponding TBUs of both devices are not independent: the absolute phase difference $|\Delta\varphi_{0}| = |\varphi_{0,2} - \varphi_{0,1}|$, wrapped to $[0,\pi]$, has a mean of 0.53~rad (Fig.~\ref{fig:4}b). This is well below the mean of $\pi/2$ expected if both devices had uncorrelated passive phases. Therefore, the two processors share a common phase profile, inherited from the design and the fabrication process, on top of which each device presents moderate individual deviations.

To test whether a nominally identical device could emulate the target, we modified the active phase settings of the first device by injecting a fraction $\alpha$ of the measured passive phase difference relative to the second device, $\Delta\varphi_{0}$, across 100 random challenges, yielding an effective total phase of $\varphi_{\text{total},1} = \varphi_{0,1} + \alpha\,\Delta\varphi_{0} + \varphi_{\text{active}}$. Setting $\alpha = 0$ corresponds to testing uncompensated, independent passive phase profiles across both devices under identical dynamic challenges. Conversely, setting $\alpha = 1$ fully compensates for the passive mismatch, enforcing identical total phase distributions ($\varphi_{\text{total},1} = \varphi_{\text{total},2}$) in both processors.

After binarization, the fractional HD was evaluated between the key sets generated by the target device and the emulating processor, as illustrated in Figure~\ref{fig:4}c. The HD decreases smoothly from 44.5\% in the uncompensated case to 29.1\% under full compensation, which is still about ten times the intra-chip HD. The full distributions for $\alpha = 0$, 0.5 and 1 highlighted by the shaded bands in Fig.~\ref{fig:4}c correspond to the HD distributions shown in Fig.~\ref{fig:4}d to f. As the passive phase difference is compensated, the whole distribution shifts progressively towards lower HD and becomes narrower, with the standard deviation decreasing from 11\% to 9\%. Passive phases therefore account for only about one third of the inter-device difference, while approximately two-thirds originates from other fabrication tolerances, such as coupler splitting ratios, optical losses, and photodetector responsivities.

Uniqueness is therefore governed by the interplay of multiple independent physical parameters rather than any single dominant factor. As a result, an adversary equipped with a nominally identical device and full characterization of the target's passive phase profile still cannot reproduce its optical responses by passive phase compensation alone.

Figure~\ref{fig:4}d depicts the distribution of the inter-chip HD over all 28 input ports. The resulting mean value of 44.4\% closely approaches the HD obtained when evaluating distinct challenges on a single device, demonstrating that separate processors respond to identical challenges nearly as independently as to entirely different ones. The deviation from the ideal 50\% Hamming distance is attributed not to inter-device correlations, but to a static structural bias inherent to the mesh layout. Because this layout constraint cannot be fully compensated by the routing topology, certain output ports systematically receive lower optical power across both devices.

\subsection{Key space expansion}

\begin{figure}[t]
\centering
\includegraphics[width=\linewidth]{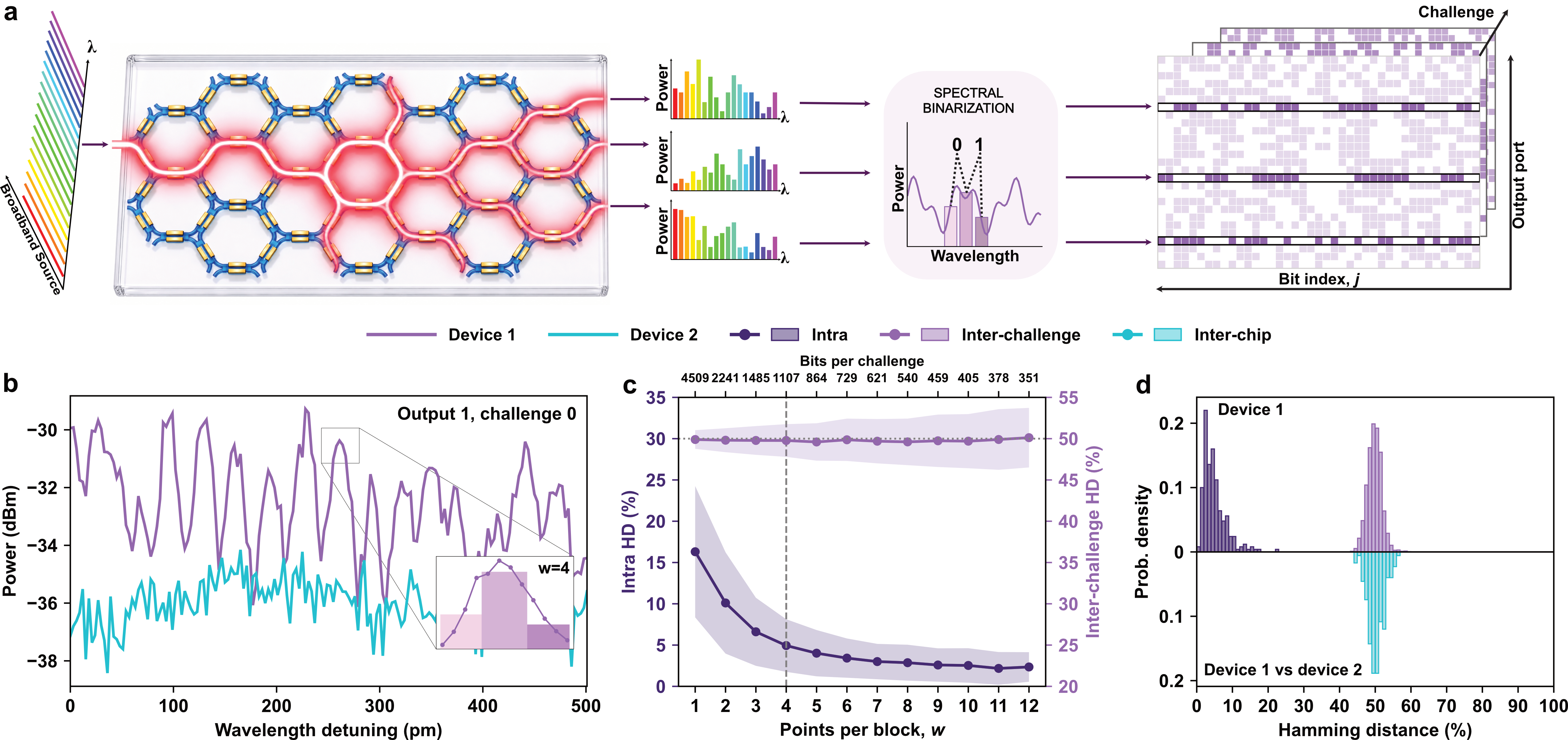}
\caption{\textbf{Key space expansion through spectral binarization.} (a) Concept: a broadband characterization of each output port, binarized by comparing adjacent wavelength blocks of the same port, extends the response of each challenge to $27 \times 41$ bits. (b) Spectral response of one output port to the same challenge in the first and second device. (c) Mean inter-challenge and intra-chip HD as a function of the number of wavelength points per block, $w$; the top axis indicates the resulting number of bits per challenge and the dashed line marks $w = 4$. (d) Distributions of the intra-chip, inter-challenge and inter-chip HD for $w = 4$ (1107 bits per challenge).}
\label{fig:5}
\end{figure}

Finally, we expanded the cryptographic key space of a single photonic device by combining our entropy-optimized mesh topology with wavelength spectrum responses, as shown in Fig.~\ref{fig:5}a. Within the mesh, applied challenge configurations induce distinct interference patterns across paths of varying lengths, yielding unique spectral responses at every output port with fine features governed jointly by the active phase settings and device-specific fabrication variations. Rather than generating a single key per input by evaluating the collective output signal as in the single wavelength operation, each output port now provides its own characteristic response to the challenge, drastically increasing the number of bits per challenge. Furthermore, comparing wavelength responses within each port rather than spatial power levels across ports eliminates static layout-induced power patterns, thereby driving the inter-chip HD closer to the theoretical 50\% ideal.

For each challenge, we measured the optical power of the 27 monitored output ports at $M = 168$ wavelengths between 1550.000 and 1550.501~nm, with a step of 3~pm. Each output port is then binarized independently along the wavelength response (Fig.~\ref{fig:5}a). Its spectrum is divided into consecutive blocks of $w$ points with mean power $\overline{P}(j)$, where $j$ is the block index, and each bit compares two adjacent blocks of the same port, such that $b_{j} = 1$ if $\overline{P}(j) > \overline{P}(j + 1)$ and $b_{j} = 0$ otherwise. Because every comparison involves a single output port, any factor that depends only on the port, including the port-dependent power pattern identified in the previous section, cancels out by construction. Each output port then provides $\lfloor M/w \rfloor - 1$ bits, and each challenge yields $27 \times (\lfloor M/w \rfloor - 1)$ bits, compared with 26 bits for the spatial operator.

The spectral operator, however, relies on every output port carrying enough power for its spectrum to be resolved above the detection noise. This is where the Isotropic Ring is essential: by distributing the power uniformly across the perimeter, it ensures that all 27 output ports provide reliable spectral bits, whereas the random routing leaves the outputs far from the input close to the detection floor, where the comparisons are dominated by noise, as observed for the spatial operator.

Figure~\ref{fig:5}b shows the spectral response of one output port to the same challenge in both devices. Although the passive phases of the two devices differ only slightly (Fig.~\ref{fig:4}b), their spectra exhibit clearly different fine structures, which the spectral operator converts into distinct bit sequences.

The block size $w$ sets the trade-off between the number of bits and their reliability (Fig.~\ref{fig:5}c). The inter-challenge HD remains between 49.5\% and 50.2\% for all block sizes between 1 and 12 points, so does not depend on this choice. The intra-chip HD, obtained from repeated spectral sweeps of the same challenge, decreases from 16.3\% for $w = 1$ to 2.4\% for $w = 12$, since each block averages more points. For small blocks, however, consecutive bits of the same output port are positively correlated, with a correlation of 0.21 for $w = 1$, because the blocks are narrower than the spectral features of the response and part of the bits are redundant. This correlation vanishes between $w = 4$ and $w = 6$, indicating a characteristic spectral correlation width of about 12 to 18~pm (see Supplementary Note~\ref{sn:spectral}). We therefore use $w = 4$, the smallest block size for which adjacent bits are essentially uncorrelated (correlation of 0.04), which yields 41 bits per output port and 1107 bits per challenge, about 43 times (one order of magnitude) more than the spatial operator.

Figure~\ref{fig:5}d shows the resulting HD distributions for all the input ports and 25 challenges measured in both devices. The inter-configuration HD within the first device reaches 49.8\%, and the inter-chip HD for identical challenges reaches 50.0\%. The spectral operator therefore restores ideal uniqueness, and the two devices respond to the same challenge as independently as to different ones. The intra-chip HD is 4.9\% on average, with a maximum of 22.1\%, well separated from the minimum inter-chip HD of 43.0\%. Combining an entropy-optimized mesh topology with a spectral readout therefore increases the key capacity by more than one order of magnitude while achieving ideal uniqueness for all tested challenges on both devices.

\section{Discussion}

We have shown that the topology of a programmable photonic mesh, rather than the amount of fabrication disorder, limits the entropy that can be extracted from it, and that this limit can be lifted by reprogramming the optical routing alone. By neutralizing the structural bias of the mesh, the Isotropic Ring topology harvests entropy from all dynamic TBUs simultaneously, rather than from isolated sub-regions synthesized as local circuits. Operation in both single-wavelength and broadband modes provides the flexibility to tailor key length and reliability on demand. This allows for direct, high-entropy key extraction from raw optical responses without post-processing projections or hashing to enhance uniqueness; standard error correction is still required to absorb the residual intra-chip HD. Table~\ref{tab:1} summarizes state-of-the-art integrated programmable photonic PUFs demonstrated to date.

\begin{table}[t]
\centering
\caption{Comparison with previously reported integrated programmable photonic PUFs.}
\label{tab:1}
\footnotesize
\setlength{\tabcolsep}{3.5pt}
\renewcommand{\arraystretch}{1.25}
\begin{tabularx}{\linewidth}{>{\raggedright\arraybackslash}X >{\raggedright\arraybackslash}p{2.75cm} >{\raggedright\arraybackslash}p{1.45cm} >{\raggedright\arraybackslash}p{2.1cm} >{\centering\arraybackslash}p{1.45cm} >{\centering\arraybackslash}p{1.45cm} >{\centering\arraybackslash}p{1.0cm} >{\centering\arraybackslash}p{1.0cm}}
\toprule
\textbf{Platform} & \textbf{Readout} & \textbf{Entropy extraction} & \textbf{Post-processing} & \textbf{Bits/\newline challenge} & \textbf{Reconfig. elements} & \textbf{Intra-HD} & \textbf{Inter-HD} \\
\midrule
SOI rectangular MZI mesh \cite{zhu2026} & All-port power readout & Full-Mesh & None & 8 & 40 TBUs & 2.6\% & 57.7\% \\
SOI hexagonal MZI mesh \cite{sarantoglou2025} & Input-Output recurrent spectrum slicing & Localized & Random projection & 256 & 8 TBUs (7 fixed) & $\approx$5\% & $\approx$45\% \\
SOI hexagonal MZI mesh \cite{tzekas2026} & Input-Output multi-resonant structures & Localized & Gabor filtering + PCA & Not specified & 13 TBUs (11 fixed) & 9.5\% & 48.4\% \\
SiN feedforward MZI mesh \cite{sarantoglou2026} & All-port power readout & Full-Mesh & Random projection + XOR & 256 & 5 TBUs + 5 PS & 2.6\% & 46.3\% \\
\midrule
\textbf{SOI hexagonal MZI mesh (this work)} & \textbf{All-port spectral readout} & \textbf{Full-Mesh} & \textbf{None} & \textbf{1107} & \textbf{72 TBUs (8-12 fixed)} & \textbf{4.9\%} & \textbf{49.8\%} \\
\bottomrule
\end{tabularx}
\end{table}

Our approach achieves comparable reliability and near-ideal uniqueness while generating two orders of magnitude more raw bits than previous photonic mesh implementations without post-processing. Furthermore, without any post-processing, it yields more than four times the key length achieved by architectures that rely on postprocessing techniques, which remain fully compatible with our scheme and can still be applied to expand the total bit count. Moreover, our architecture incorporates the largest number of reconfigurable elements, offering more than 60 programmable TBUs for implementing challenges, which directly translates into a higher CRP dimensionality.

Beyond establishing routing as a tool to counteract layout-induced bias, these results pave the way for investigating scaling dynamics, footprint-to-entropy scaling laws, and next-generation topologies engineered to minimize physical bias. Promising future directions include geometrically symmetric mesh geometries that neutralize rectangular boundary effects (e.g., circular layouts), alongside central light injection schemes using grating couplers. The latter would remove the need to allocate TBUs to route light to the core, further increasing the net entropy harvested from the processor.

Because the approach relies solely on reprogramming commercially available processors, it can be applied directly to existing hardware. By delivering the highest raw key density reported for mesh-based PUFs together with near-ideal uniqueness, this work positions programmable photonic processors as reconfigurable hardware roots of trust for optical communication, sensing and computing systems, in which the same chip that processes information can also secure it.

\section{Methods}

\subsection{Experimental setup}

Experimental measurements were performed using two iPronics Smartlight processors and an external tunable laser source EXFO T100S-HP. The laser was operated in CW mode at 1550~nm for single-wavelength measurements, and in swept-wavelength mode for spectral measurements, maintaining an output power of 11~dBm in both cases. Signal detection was achieved using Smartlight's integrated photodetectors, which exhibit a noise floor of approximately $-40$~dBm. To establish the active phase configurations, we used the calibration data provided for each processor, which maps the applied drive current to the induced optical phase shift across each TBU.

\subsection{Isotropic Ring topology routing}

Under the Isotropic Ring topology, routing proceeds from every input port to the spatial center of the mesh through a fixed path of TBUs set to the cross or bar state. Because of the rectangular layout geometry, the length of this path depends on the position of the input port, from 2 TBUs for the ports closest to the center to 6 TBUs for the farthest ones. At the spatial center, six TBUs synthesize a closed hexagonal loop. Equal power distribution across all external output ports is enforced by setting the power coupling coefficient $k_{j}$ of the $j$-th TBU as $k_{j} = 1/(7 - j)$. The routing path and the ring therefore use between 8 and 12 TBUs with fixed settings, and the remaining 60 to 64 TBUs, depending on the input port, are the dynamic TBUs whose phases are set by the challenge.

\subsection{Shannon entropy}

The Shannon entropy of the normalized power vector quantifies the spatial bias of the optical routing \cite{shannon1948}. High entropy corresponds to an ideal, homogeneously coupled beam splitter with optimal power mixing, whereas low entropy reflects spatial energy localization across a small fraction of output ports, characteristic of selective filtering. For a given challenge the entropy $H(\mathbf{P})$ of the normalized response vector is given by:
\begin{equation}
H(\mathbf{P}) = - \sum_{n = 1}^{N} P_{n}\log_{2}(P_{n}),
\end{equation}
where $P_{n}$ represents the power at each one of the $N$ outputs. Under the ideal condition of equal power distribution across all output ports ($P_{n} = 1/N$), the Shannon entropy reaches its theoretical maximum of $H(\mathbf{P})_{\max} = \log_{2}(N)$ bits.

\subsection{Hamming distance}

To quantitatively evaluate the stability and uniqueness of the extracted $L$-bit binary keys, we use the normalized Hamming distance. For two binary keys $K_{A},K_{B} \in \{0,1\}^{L}$, the HD is defined as the normalized bitwise exclusive-OR (XOR) sum:
\begin{equation}
\mathrm{HD}(K_{A},K_{B}) = \frac{1}{L}\sum_{n = 1}^{L} \left( K_{A,n} \oplus K_{B,n} \right),
\end{equation}
where $\oplus$ denotes the XOR operator. While intra-HD measures key reproducibility and system stability between keys generated under identical nominal challenge conditions, inter-HD quantifies key uniqueness and unpredictability across keys derived from distinct phase configurations or independent input challenges. Consequently, the ideal target is 0\% for intra-HD and 50\% for inter-HD, where any positive or negative deviation from 50\% indicates reduced key randomness.

\section*{Data availability}
The data supporting the findings of this study are available from the corresponding authors upon reasonable request.

\section*{Code availability}
The code used to analyse the data and generate the figures is available from the corresponding authors upon reasonable request.

\section*{Acknowledgements}
This work was supported by ERC-ADG-2022-101097092 ANBIT, ERC-POC-2025-1 101241773 TRANSBIT MESH, GVA PROMETEO 2021/015 research excellency award, Fundación BBVA Programa de Investigación Fundamentos 2024 API project, Ministerio de Ciencia y Universidades Plan Complementario de Comunicación Cuántica projects QUANTUMABLE-1 and QUANTUMABLE-2, and HUB de Comunicaciones Cuánticas.

\section*{Author contributions}
J.C. conceived the concept and supervised the overall project. O.A developed the control software and carried out the experimental demonstration of the application. PM participated in the discussion of the results and provided comments on the paper. O.A, P.M and J.C. wrote the original draft of the paper. All authors reviewed and approved the final version.

\section*{Competing interests}
The authors declare that they have no conflicts of interest.

\renewcommand{\refname}{References}

\clearpage
\setcounter{figure}{0}
\setcounter{table}{0}
\setcounter{equation}{0}
\setcounter{section}{0}
\renewcommand{\figurename}{Supplementary Figure}
\renewcommand{\tablename}{Supplementary Table}
\renewcommand{\theequation}{S\arabic{equation}}
\renewcommand{\thesection}{\arabic{section}}
\titleformat{\section}{\normalfont\large\bfseries}{Supplementary Note~\thesection}{0.8em}{}

\begin{center}
{\LARGE\bfseries Supplementary Information\par}
\vspace{0.4em}
{\large Photonic processors that carry their own unclonable fingerprint\par}
\end{center}
\vspace{0.6em}

\section{Port mapping and routing}
\label{sn:ports}

The SmartLight processor provides 40 optical ports along the perimeter of the mesh, of which 28 are accessible in our setup (physical ports 0 to 21 and 34 to 39). In the main text figures, these ports are renumbered consecutively from 0 to 27, following their position along the perimeter. For each input channel, the remaining 27 ports serve as monitored outputs, maintaining their sequential perimeter order while omitting the input port itself. This arrangement defines the spatial adjacency utilized by the spatial binarization operator.

\begin{figure}[h]
\centering
\includegraphics[width=0.6\linewidth]{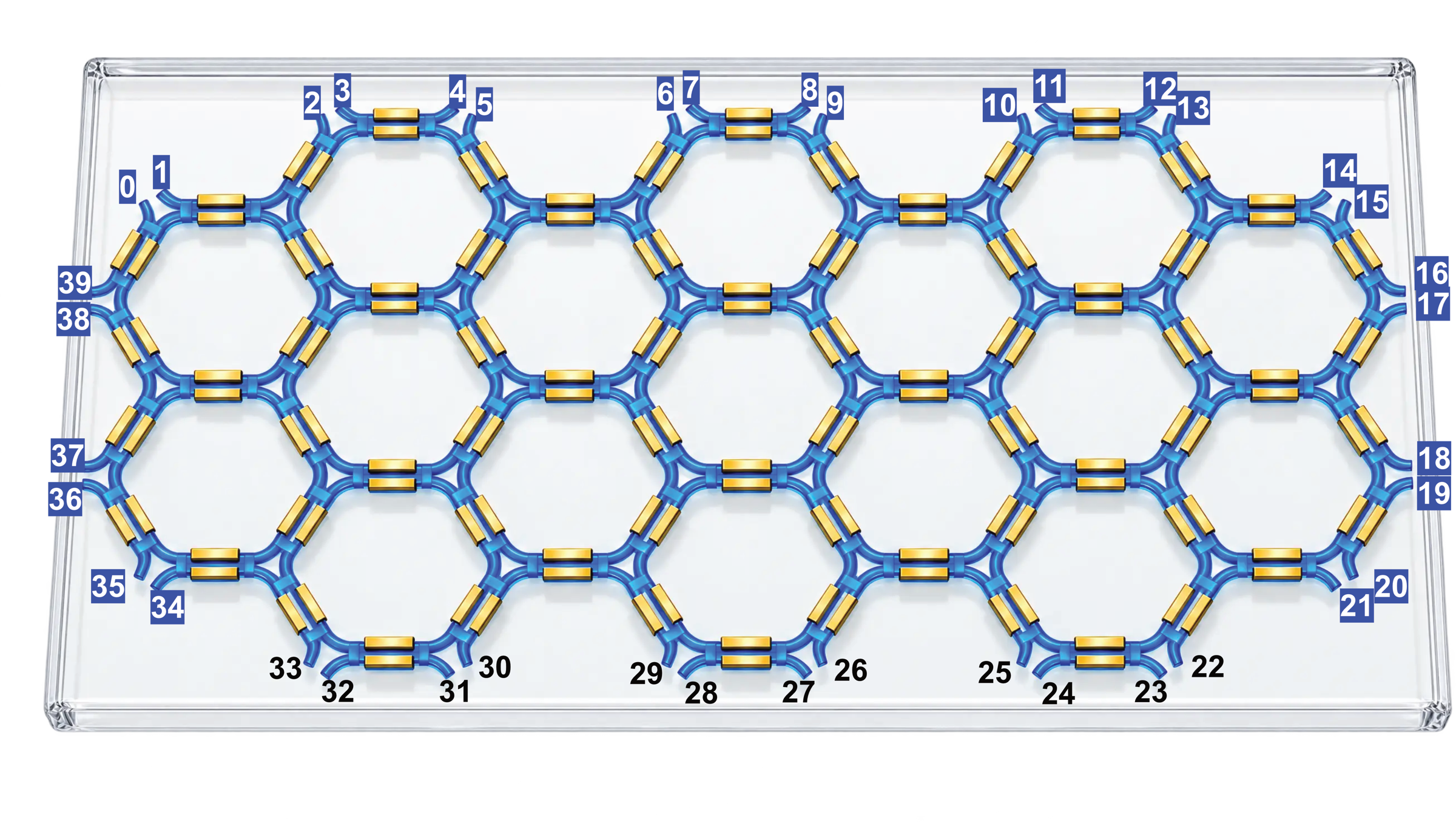}
\caption{Port numbering of the programmable photonic processor, shown on a 3D model of the hexagonal mesh. The 40 optical ports along the perimeter are labeled with their physical index. The 28 ports accessible in our setup, physical ports 0 to 21 and 34 to 39, are highlighted in blue.}
\label{fig:ports}
\end{figure}

In the Isotropic Ring topology, the light from each input port is routed to the central hexagonal ring through a fixed path of TBUs set to the cross or bar state. Supplementary Table~\ref{tab:routing} lists, for each input port, the number of TBUs in the access path, the total number of TBUs with fixed settings, and the number of dynamic TBUs that receive the challenge phases.

\begin{table}[h]
\centering
\caption{Routing of the Isotropic Ring topology for each accessible input port. The fixed TBUs include the access path and the six TBUs of the ring. The dynamic TBUs are the remaining TBUs of the 72-TBU mesh, whose phases are set by the challenge.}
\label{tab:routing}
\small
\begin{tabular}{ccccc}
\toprule
Input port (figure index) & Physical port & Access path TBUs & Fixed TBUs & Dynamic TBUs \\
\midrule
0 & 0 & 6 & 12 & 60 \\
1 & 1 & 6 & 12 & 60 \\
2 & 2 & 5 & 11 & 61 \\
3 & 3 & 5 & 11 & 61 \\
4 & 4 & 6 & 12 & 60 \\
5 & 5 & 4 & 10 & 62 \\
6 & 6 & 2 & 8 & 64 \\
7 & 7 & 3 & 9 & 63 \\
8 & 8 & 3 & 9 & 63 \\
9 & 9 & 2 & 8 & 64 \\
10 & 10 & 3 & 9 & 63 \\
11 & 11 & 6 & 12 & 60 \\
12 & 12 & 5 & 11 & 61 \\
13 & 13 & 5 & 11 & 61 \\
14 to 21 & 14 to 21 & 6 & 12 & 60 \\
22 to 27 & 34 to 39 & 6 & 12 & 60 \\
\bottomrule
\end{tabular}
\end{table}

\clearpage
\section{Benchmark of binarization methods}
\label{sn:binarization}

The spatial differential operator used in the main text was selected after comparing five binarization strategies. All of them are applied to the same measured power vector $\mathbf{P} = (P_1, \dots, P_N)$ of the $N = 27$ monitored outputs of one input port, expressed in dB and ordered by their position along the perimeter of the mesh (Supplementary Note~\ref{sn:ports}).

\paragraph{Threshold at the vector mean.} Each output is compared with the mean power of the same measurement,
\begin{equation}
  b_n = \begin{cases} 1 & \text{if } P_n > \bar{P} \\ 0 & \text{otherwise} \end{cases}, \qquad \bar{P} = \frac{1}{N}\sum_{n=1}^{N} P_n, \qquad n = 1, \dots, N,
\end{equation}
which gives 27 bits per configuration.

\paragraph{Threshold at the vector median.} The same, with the threshold set at the median of the 27 powers of the measurement, $\tilde{P} = \operatorname{median}(P_1, \dots, P_N)$, which gives 27 bits.

\paragraph{Spatial differential.} Each output is compared with the next one along the perimeter, as in the main text,
\begin{equation}
  b_n = \begin{cases} 1 & \text{if } P_n > P_{n+1} \\ 0 & \text{otherwise} \end{cases}, \qquad n = 1, \dots, N - 1,
\end{equation}
which gives 26 bits.

\paragraph{Binary rank.} Each output is replaced by its rank in the ordered power vector,
\begin{equation}
  r_n = \left|\{\, m : P_m > P_n \,\}\right|, \qquad r_n \in \{0, \dots, N - 1\},
\end{equation}
so that the brightest output has rank 0. Each rank is written in natural binary with 5 bits, the minimum for 27 values, which gives $5N = 135$ bits.

\paragraph{Lehmer code.} The ordering of the 27 powers is a permutation, which is encoded by its Lehmer code,
\begin{equation}
  L_n = \left|\{\, m > n : P_m < P_n \,\}\right|, \qquad L_n \in \{0, \dots, N - n\}, \qquad n = 1, \dots, N - 1.
\end{equation}
Each element $L_n$ counts the number of subsequent ports in the perimeter sequence ($m > n$) that exhibit a strictly lower optical power than port $n$. Digits are written with the minimum number of bits for their range, $\lceil \log_2(N - n + 1) \rceil$, using a Gray code, $g_n = L_n \oplus \lfloor L_n / 2 \rfloor$, so that consecutive values of the digits differ in a single bit. This gives $\sum_{n=1}^{N-1} \lceil \log_2(N - n + 1) \rceil = 104$ bits.

All binarization methods depend solely on the relative output powers within a single measurement, making them inherently insensitive to common-mode fluctuations such as variations in laser power or input coupling efficiency. These strategies were evaluated by analyzing standard figures of merit of the generated cryptographic keys, as well as the intra-chip and inter-configuration Hamming distances. For a method producing $K$ bits per configuration, let $b_k^{(c)}$ be bit $k$ of the response to configuration $c$, for $C = 500$ configurations of one input port. The uniformity of a response is the fraction of ones, $\frac{1}{K}\sum_{k} b_k^{(c)}$, averaged over the configurations, with an expected value of 50\%. The bit-aliasing of position $k$ is the fraction of configurations in which the bit takes the value 1, $\mathrm{BA}_k=\frac{1}{C}\sum_{c=1}^{C} b_k^{(c)}$, computed across configurations of a single device, whose ideal value is 50\%. Supplementary Table~\ref{tab:binarization} reports the mean deviation from this ideal, $|\mathrm{BA}_k - 50\%|$, averaged over the bit positions and the 28 input ports, whose ideal value is 0.

Specifically, the inter-configuration HD was computed across all pairwise combinations of the 500 configurations per input port. For the intra-chip HD, 100 configurations of each topology were measured over 10 consecutive rounds per port, interleaving the Isotropic Ring and random configurations with a 0.5~s settling time after programming each state. The intra-chip HD was then calculated between consecutive measurement rounds of the same configuration for each binarization method. Supplementary Table~\ref{tab:binarization} summarizes the results.

\begin{table}[h]
\centering
\caption{Comparison of binarization methods for both topologies. Intra: intra-chip HD between consecutive readings. Inter: inter-configuration HD. Uniformity: mean fraction of ones per response. $|\mathrm{BA}-50|$: mean deviation of the bit-aliasing from 50\% over the bit positions (ideal value 0).}
\label{tab:binarization}
\small
\begin{tabular}{lccccc}
\toprule
Method & Bits & Intra (\%) & Inter (\%) & Uniformity (\%) & $|\mathrm{BA}-50|$ (\%) \\
\midrule
\multicolumn{6}{l}{\textit{Isotropic Ring}} \\
Threshold at the vector mean & 27 & 2.3 & 44.2 & 48.0 & 13.6 \\
Threshold at the vector median & 27 & 2.6 & 44.4 & 48.1 & 13.5 \\
Spatial differential ($d = 1$) & 26 & 3.5 & 45.6 & 50.8 & 10.9 \\
Binary rank & 135 & 16.5 & 47.8 & 44.4 & 7.0 \\
Lehmer code (Gray) & 104 & 8.1 & 44.8 & 47.4 & 11.7 \\
\midrule
\multicolumn{6}{l}{\textit{Random}} \\
Threshold at the vector mean & 27 & 1.7 & 30.1 & 39.9 & 27.7 \\
Threshold at the vector median & 27 & 3.5 & 32.0 & 48.1 & 27.0 \\
Spatial differential ($d = 1$) & 26 & 6.4 & 41.5 & 50.9 & 16.2 \\
Binary rank & 135 & 18.3 & 45.1 & 44.4 & 9.5 \\
Lehmer code (Gray) & 104 & 10.2 & 39.7 & 47.1 & 16.7 \\
\bottomrule
\end{tabular}
\end{table}

The binary rank and the Lehmer code provide more bits per configuration, but they are much less reliable (intra-chip HD of 16.5\% and 8.1\% for the Isotropic Ring), because a single swap between two outputs of similar power changes several bits. Among the remaining methods, we selected the spatial differential operator, as it achieves the best overall performance in terms of inter-HD, uniformity and bit-aliasing, at the expense of a slight increase in intra-HD and a one-bit reduction in key length. It is worth noting that this stage offers considerable design flexibility and depends heavily on the target robustness and key length; therefore, alternative binarization strategies may be adopted depending on the specific application.

\clearpage
\section{Spectral block size}
\label{sn:spectral}

The spectral response of each output port was measured at $M = 168$ wavelengths between 1550.000 and 1550.501~nm, with a step of 3~pm. The spectrum is divided into consecutive, non-overlapping blocks of $w$ points, and each bit compares the mean power of two consecutive blocks of the same port. The block size sets the trade-off between the number of bits and their reliability. To quantify the redundancy between bits, we computed the Pearson correlation coefficient between consecutive bits $b_j$ and $b_{j+1}$ of the same output port, challenge and sweep.

For $w = 1$, consecutive bits are positively correlated (0.21), because the blocks are narrower than the spectral features of the response. The correlation decreases with $w$ and falls below 0.05 for $w = 4$, which corresponds to blocks of 12~pm. For larger blocks the correlation becomes negative, because consecutive bits share their central block and the noise of that block affects both bits in opposite directions. We therefore use $w = 4$, the smallest block size for which consecutive bits are essentially uncorrelated while offering long key generation.

\begin{figure}[h]
\centering
\includegraphics[width=0.6\linewidth]{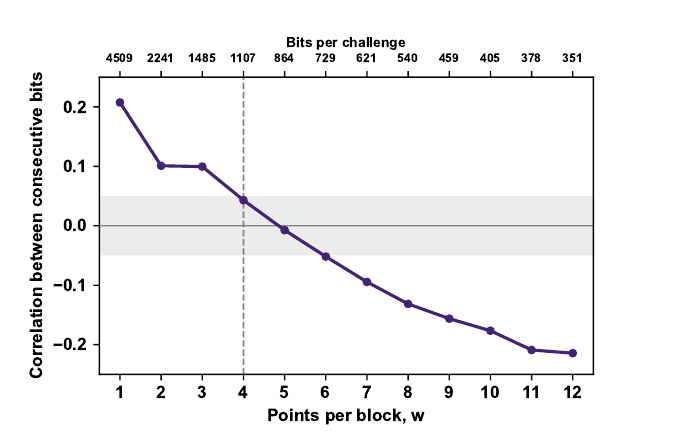}
\caption{Correlation between consecutive spectral bits of the same output port as a function of the block size $w$. The top axis gives the resulting number of bits per challenge, the shaded band marks correlations below 0.05 in magnitude, and the dashed line marks $w = 4$.}
\label{fig:corrw}
\end{figure}

\end{document}